\documentclass[aps,prl,twocolumn,superscriptaddress,showpacs]{revtex4}

\usepackage{longtable}
\usepackage{graphicx}
\usepackage{dcolumn}
\usepackage{bm}
\usepackage{amsmath}
\usepackage[psamsfonts]{amssymb}

\newcommand{\PCCO}{Pr$_{2-x}$Ce$_x$CuO$_{4-\delta}$}

\newcommand{\etal}{\emph{et al.}}

\begin{document}

\title{Single superconducting energy scale in electron-doped cuprate superconductor Pr$_{2-x}$Ce$_x$CuO$_{4-
\delta}$}


\author{I. Diamant}
\email[]{diamanti@post.tau.ac.il}\affiliation{Raymond and Beverly Sackler School of Physics and Astronomy, Tel-Aviv
University, Tel Aviv, 69978, Israel}
\author{R.L. Greene}
\affiliation{Center for nanophysics and advanced materials, Physics Department, University of Maryland, College Park,
Maryland 20743, USA}
\author{Y. Dagan}
\affiliation{Raymond and Beverly Sackler School of Physics and Astronomy, Tel-Aviv University, Tel Aviv, 69978,
Israel}


\date{\today}

\begin{abstract}
The tunneling spectra of the electron-doped cuprate \PCCO~ as a
function of doping and temperature is reported. We find that the
superconducting gap, $\Delta$, shows a BCS-like temperature
dependence even for extremely low carrier concentrations (studied
here for the first time). Moreover, $\Delta$ follows the doping
dependence of T$_c$, in strong contrast with tunneling studies of
hole-doped cuprates. From our results we conclude that there is a
single superconducting energy scale in the electron-doped
cuprates.

\end{abstract}


\maketitle


In his pioneering experiment Giaever showed that when a simple
metal and a classical superconductor are connected through an
insulating barrier the tunneling conductance is proportional to
the density of states in the superconducting electrode. He used
this method to directly measure the energy gap, $\Delta$, in
various superconductors.\cite{GiaeverRMP} Giaever found that
$\Delta$ drops to zero at the critical temperature T$_{c}$.  In
addition, for various materials $\Delta/kT_{c}$ is approximately
constant as earlier predicted by Bardeen, Cooper, and Schrieffer
(BCS).  Their theory predicts a single energy scale that is both
related to the onset of single particle excitations ($\Delta$) and
the temperature, T$_c$, at which coherence is destroyed.\cite{BCS}
\par

In the hole-doped high T$_{c}$ cuprates Renner
\etal\cite{Rennerandfisher} showed that the spectra obtained by
scanning tunneling spectroscopy exhibit no special temperature
dependence at the temperature where macroscopic superconductivity,
i.e. vanishing resistance and Meissner effect, ceases to exist.
This behavior was interpreted as a signature of the pseudogap
state, which can be detected in a variety of
experiments.\cite{Timuskandstat} The nature of this pseudogap and
its relation to superconductivity are still a puzzle. It has been
suggested that the pseudogap is precursor
superconductivity,\cite{Chen_Levin_PG_dueto_PF_Pairs} a competing
order parameter\cite{Loramtallon} or a phenomenon related to the
range of the antiferromagnetic
interactions.\cite{friedel_PGdue_toAFM}
\par

Deutscher \cite{Deutscher_nature} has pointed out that for the
hole-doped cuprates there are two energy scales that merge
together at high doping levels: the lower one, which follows
T$_c$, is the phase coherent energy scale, probed by
Andreev-Saint-James reflections. The higher energy scale is
related to single particle excitations. It increases monotonically
with decreasing doping. More recent contributions have confirmed
Deutscher's observation of two energy scales for the hole-doped
cuprates. \cite{LeeBCSnearnode, Tacontwogaps, Hufner_twogaps}
\par
The electron-doped and the hole-doped cuprates share many
structural and electronic properties:\cite{FournierBook} they both
comprise of copper oxygen planes, were \textit{d}-wave
superconductivity takes place.\cite{TsueiRMP} The parent compounds
are antiferromagnetic insulators, which become superconducting
upon adding charge carriers (doping) in a dome shaped region in
the temperature-doping phase diagram.  For electron-doped cuprates
the Fermi surface evolves from small electron pockets in the
underdoped regime into a large hole-like Fermi surface on the
overdoped side.\cite{armitageprldoping, matsui_fullFS} For hole
doped cuprates possible evidence for electron pockets were found
in underdoped YBa$_{2}$Cu$_{3}$O$_{6.5}$ in quantum oscillations
measurements. \cite{LeBoeufYBCO} Similar measurements on the
overdoped side were interpreted in terms of large hole-like fermi
surface.\cite{Vignolle_QOSinTL}
\par
On the other hand, there are
several differences between the two types of cuprates: while for
hole-doped the antiferromagnetic phase dissapears rapidly with
doping it is relatively extended on the electron doped side,
possibly persisting into the superconducting dome.\cite{LukeAFM,
fujita:147003} The temperature dependence of the resistivity well
above T$_c$ is very different for hole and electron-doped
cuprates.\cite{FournierBook} Finally, possible existence of higher
harmonics in the order parameter for electron-doped
superconductors has been reported by several
groups.\cite{matsuiNMD, blumbergNmd, DaganDirtySC}
\par
For hole-doped cuprates the pseudogap and the superconducting gap
coexist both in doping and momentum space, they intermix for many
spectroscopic probes (an exception is Andreev-Saint-James
reflections that are sensitive only to the superconducting state). The superconducting state may possibly be obscured by
the pseudogap for underdoped samples for most momentum directions.
By contrast, the superconducting gap is not obscured by the
pseudogap for electron-doped cuprates. Therefore, the
superconducting gap in electron-doped cuprates can be measured
directly by tunneling spectroscopy.
\par
We make use of the absence of a pseudogap phase in the
electron-doped cuprates to directly measure the full doping and
temperature dependence of the superconducting gap. We show that
for these compounds there is a single superconducting energy
scale, $\Delta$, which follows the same doping dependence as
T$_{c}$ for the entire phase diagram, even for the heavily
underdoped region (samples with T$_{c}$ as low as 6K). Assuming
that the two types of cuprates share the same mechanism
responsible for superconductivity, our results may imply that the
pseudogap state in the hole-doped cuprates is a competing order to
the superconducting one.
\par
We fabricated superconductor/insulator/superconductor (SIS)
junctions using \PCCO~(PCCO) and lead as described elsewhere.
\cite{daganQazilbashtunneling,DaganDirtySC} The advantage of
planar tunnel junctions is using the ability to measure the SIS
tunneling conductance at various temperatures and magnetic fields
without changing the properties of the junction. This is in strong
contrast with scanning tunneling microscopy measurements where a
change in temperature or magnetic field may result in junction
resistance variation due to its exponential dependence on
tip-sample distance. At high magnetic fields, $\mu_0$H=14T,
superconductivity in the PCCO is muted and the normal state is
revealed. This enables us to normalize the data as was done by
Giaever.\cite{GiaeverRMP} This eliminates spurious barrier
and normal state effects. This procedure is impossible for
hole-doped cuprates due to its inaccessible upper critical field.
\par
\begin{figure}
\includegraphics[width=1\hsize]{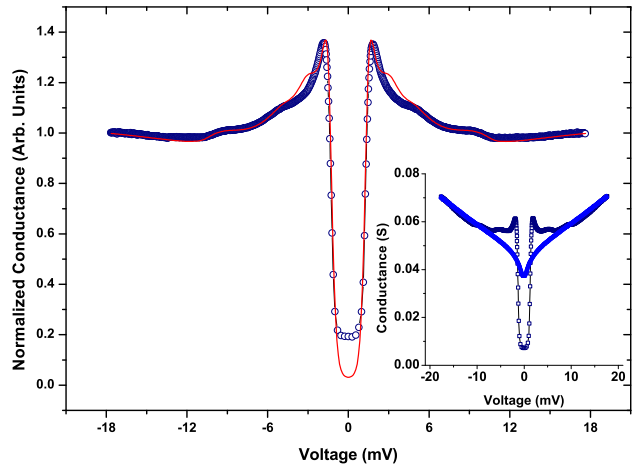}
\caption {Tunneling spectra of heavily underdoped PCCO x=0.125 at
T=2K. The circles are the normalized data and the
solid line is a fit (see text for details) with T$_{c}$=
6.5$\pm$0.5K. We obtained $\Delta_{(max)}$=1.5$\pm $0.3
meV, $\Gamma$=0.85$\pm$0.05 meV, Z=8,
$\theta_{(max)}$=40$^\circ\pm$10$^\circ$,
$\mu_0$H$_{C2}$=1.5$\pm$0.5 Tesla, $2\Delta/K_BT_c$=5.4$\pm$1.1.
Insert: A plot of the differential conductance at zero magnetic
field ($\square$) and at $\mu_0$H=14 Tesla ($\vartriangle$) which is
used to normalize the data.\label{0125Tunnel}}
\end{figure}

The conductance versus voltage for a typical very underdoped
x=0.125 junction is shown in figure ~\ref{0125Tunnel}. The strong
phonon structure of the lead (at $\pm5 meV$ and at $\pm10 meV$),
and the relatively low conductance at zero bias are indicative of
the high quality of the junctions. The insert of figure
~\ref{0125Tunnel} presents the differential conductance as a
function of voltage at zero field and at an applied field of 14
Tesla. At high magnetic fields a small reduction in the zero bias
conductance is revealed. This behavior has been reported in other
tunneling measurements on electron-doped
cuprates.\cite{KleefischNSTG, BiswasNSTG, alffnature,
daganQazilbashtunneling}
\par
We fit the data using a Blonder-Tinkham-Klapwijk \cite{BTK}
model extended for anisotropic order
parameters.\cite{KashiwayaTheory} We used a modified
\textit{d}-wave gap, which was suggested for electron-doped
cuprates.\cite{blumbergNmd, ereminChubukovNMD} This modified
\textit{d}-wave gap better fits the Raman,\cite{blumbergNmd}
ARPES,\cite{matsuiNMD} and tunneling\cite{DaganDirtySC} spectra.
In this model the gap has a maxima away from the $(\pi, 0)$, at an
angle $\theta_{(max)}$. We used Z, $\Delta$, $\theta_{(max)}$ and
$\Gamma$ as free parameters, with Z being the barrier strength and
$\Gamma$ is a lifetime broadening.\cite{dyneslifetime} More
details on the fitting procedure are described
elsewhere.\cite{DaganDirtySC}
\par
We emphasize that the gap amplitude, which is the main focus of
this contribution, is independent of the details of the  order
parameter chosen for the fit. The gap amplitude is determined
predominantly by the energy at which the coherence peaks appear at
low temperatures.

\par
\begin{figure}
\includegraphics[width=1\hsize]{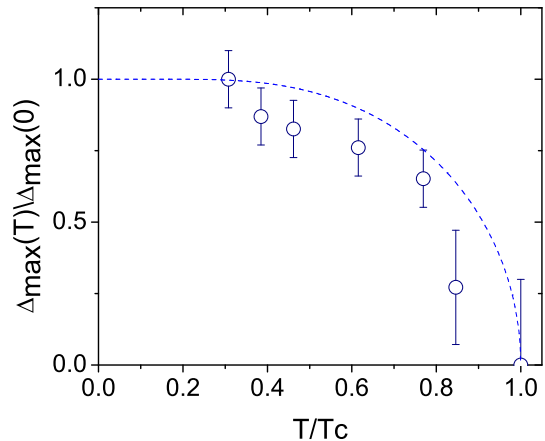}
\caption{Reduced gap amplitude of the PCCO x=0.125 plotted as a function of the
reduced temperature. The circles are the reduced gap amplitude at different reduced temperatures as found from each
fit. The dashed line is the BCS prediction.\label{DeltaT}}
\end{figure}

In figure ~\ref{DeltaT} we show the temperature dependence of the
gap maximum as found form fitting the tunneling spectra at various
temperatures for the PCCO x=0.125 sample. We point out that all fitting
parameters are determined at the lowest temperature, leaving the
gap amplitude as the only temperature dependent fitting parameter.
For comparison the BCS prediction is shown. This result is similar
to the temperature dependence reported for higher
dopings.\cite{DaganDirtySC, shanBCS}

\par
\begin{figure}
\includegraphics[width=1\hsize]{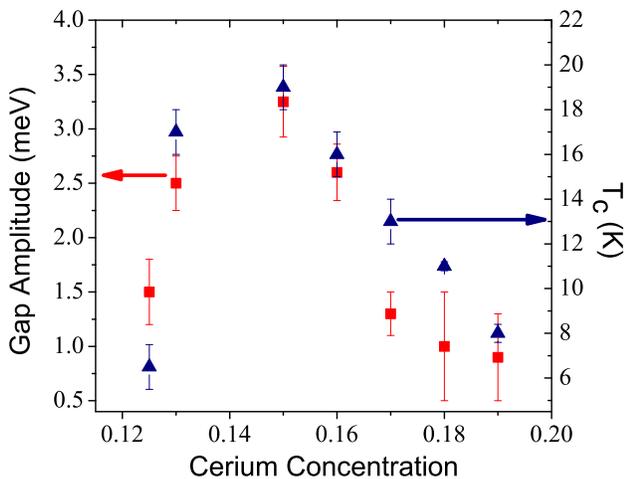}
\caption{(Color online) The obtained gap amplitude as a function of doping ($\square$).
The critical temperature as a function of doping ($\triangle$).\label{DeltaX}}
\end{figure}

In figure ~\ref{DeltaX} we present the obtained gap amplitude at
low temperatures as a function of doping. This is the main result
of this paper. We note that the gap decreases when decreasing the
doping towards the underdoped regime.

\par
Our result is in strong contrast with scanning tunneling
spectroscopy data on hole-doped cuprates.\cite{fischer_RMP} To
better understand the similarities and differences between the two
types of cuprates we shall now discuss the various gap
spectroscopies on hole-doped cuprates, and compare their findings
to our results on the electron-doped PCCO. Following
Deutscher\cite{Deutscher_nature} and the recent ARPES measurements
\cite{LeeBCSnearnode}, the various results fall into two classes:
The first class of experiments includes probes that are mostly
sensitive to the $(\pi,0)$ or anti-nodal direction. The second
class includes experiments that are sensitive to the nodal
direction $(\pi,\pi)$. Experiments that belong to the first class
such as: Raman scattering in the B$_1g$ channel, most scanning
tunneling spectroscopies and most angle resolved photoemission
spectroscopy experiments report a gap that increases with
decreasing doping on the underdoped side. This is the \textbf{k}
direction at which the pseudogap is
maximal.\cite{DingPG_has_dwavesymmetry} Scanning tunneling
spectroscopy experiments in their common configuration, i.e. the
tip perpendicular to the CuO$_2$ plane, are mostly sensitive to
the antinodal momentum direction. In this configuration, at zero
bias one can tunnel only into the nodal direction in which the gap
(and the pseudogap) is zero. As the energy is increased the
momentum cone opens up and the measurement is dominated by momenta
away from the nodal direction. For this reason the gap features in
scanning tunneling spectroscopy arise mainly from the anti-nodal
direction in momentum space. It is therefore not surprising that
for the hole-doped cuprates the gap amplitude measured by scanning
tunneling spectroscopy increases with decreasing doping on the
underdoped side\cite{fischer_RMP}, as observed for all measurements
that probe momenta along the antinodal direction. One should
therefore bear in mind that for the hole-doped cuprates a
measurement of the gap by tunneling or ARPES is not necessarily a
measurement of the superconducting order parameter.
\par
On the other hand Raman B$_2g$ channel and the slope of the
penetration depth as a function of temperature are both mostly
sensitive to the nodal direction. Such nodal sensitive
measurements show a gap that follows the doping dependence of
T$_c$ on the hole-doped side of the phase
diagram.\cite{Deutscher_nature}
\par
We can therefore conclude that for hole-doped cuprates probes
exciting single particles such as Raman scattering, Angle resolved
photoemission spectroscopy or tunneling can probe the
superconducting gap depending on their momentum selectivity, i.e.
the superconducting gap is observed for the nodal direction, while
the pseudogap dominates for the anti-nodal one. This picture is
consistent with recent ARPES measurements focusing on the nodal
region of Bi$_2$Sr$_2$CaCu$_2$O$_{8+\delta}$.\cite{LeeBCSnearnode}
Andreev-Saint-James reflections exhibit similar doping dependence
as the nodal sensitive probes, and can therefore be associated
with the second class.\cite{Deutscher_nature}
\par
Indeed, a tunneling study into the nodal direction of the
hole-doped cuprate YBa$_2$Cu$_3$O$_{7-\delta}$ showed a gap that
decreases with doping on the underdoped side.\cite{DaganEPL2} This
is consistent with the idea that when probing the nodal direction,
one is sensitive solely to the superconducting energy gap. Such
measurements that are sensitive to nodal momenta give similar
doping dependence for the gap as Andreev-Saint James reflections
that are only sensitive to the coherent state.
\par
Our results in electron-doped cuprates of an order parameter that
follows a BCS temperature and doping dependence are therefore in
line with nodal gap measurements in hole doped cuprates. This
suggests that the energy scale relevant for superconductivity
measured in our experiment by simple tunneling experiment is
related to the nodal energy scale found in hole doped cuprates.
\par
In Summary, we present tunneling spectra measurements on
Lead/Insulator/\PCCO~ over the entire doping range where
superconductivity is observed. From these spectra we extracted the
gap amplitude for each doping and at various temperatures. Our
results show a BCS like temperature dependence for the
superconducting gap even in the very underdoped regime. We show
that the gap amplitude follows the doping dependence of the
critical temperature T$_c$. This is in strong contrast with the
celebrated doping dependence of the pseudogap for the hole-doped
cuprates. Our results are therefore consistent with a single
superconducting energy scale.

We can further assume that the hole and electron doped cuprates share the same mechanism for superconductivity. In addition, one can note that for hole-doped cuprates the gap probed by Andreev-Saint-James reflections or by spectroscopy sensitive to the nodal direction follows the same doping dependence as our tunneling gap. We therefore conclude that for hole-doped  cuprates the nodal gap is related to superconductivity, while the pseudogap may be a competing order.

\begin{acknowledgments}
We are indebted to Guy Deutscher, Shay Hacohen-Gourgy for fruitful
discussions, to M. Karpovski for evaporating Lead electrodes. This
research was partially supported by the Binational Science
Foundation grant number 2006385, the Israel Science Foundation
grant number 1421/08 and by the Wolfson Family Charitable Trust.
RLG is partially supported by the NSF DMR 0653535.
\end{acknowledgments}

\bibliographystyle{apsrev}
\bibliography{Gapinpcco}
\end{document}